\pdfoutput=1

\documentclass[preprint2]{aastex62}
\usepackage{booktabs}
\usepackage{colortbl}
\usepackage{graphicx}
\usepackage[caption=false]{subfig}
\usepackage[figuresleft]{rotating}
\let\splitbox\relax
\usepackage{adjustbox}
\PassOptionsToPackage{hyphens,spaces,obeyspaces}{url}
\usepackage{amsmath,multirow,soul}
\usepackage[T1]{fontenc}
\usepackage{ textcomp }
\usepackage{gensymb}
\usepackage{hyperref}

\RequirePackage{lineno}
\accepted{May 26, 2021}
\submitjournal{ApJ}

\definecolor{redc}{RGB}{138, 22, 3}

\shorttitle{An Archival Search for NS Mergers in GWs and VHE Gamma Rays}
\shortauthors{VERITAS et al.}
\begin{document}
% \linenumbers
\title{An Archival Search for Neutron-Star Mergers in Gravitational Waves and Very-High-Energy Gamma Rays}

\correspondingauthor{C.~B.~Adams}
\email{ca2762@columbia.edu}
%%%%%%%%%%%%%%%%%%%%%%%%%%%%%%%%%%%%%%%%%%%%%%%%
%%%%%%%%%%%%%%%%%%%%%%%%%%%%%%%%%%%%%%%%%%%%%%%%
\author{C.~B.~Adams}\affiliation{Physics Department, Columbia University, New York, NY 10027, USA}
\author{W.~Benbow}\affiliation{Center for Astrophysics $|$ Harvard \& Smithsonian, Cambridge, MA 02138, USA}
\author{A.~Brill}\affiliation{Physics Department, Columbia University, New York, NY 10027, USA}
\author{J.~H.~Buckley}\affiliation{Department of Physics, Washington University, St. Louis, MO 63130, USA}
\author{M.~Capasso}\affiliation{Department of Physics and Astronomy, Barnard College, Columbia University, NY 10027, USA}
\author{J.~L.~Christiansen}\affiliation{Physics Department, California Polytechnic State University, San Luis Obispo, CA 94307, USA}
\author{A.~J.~Chromey}\affiliation{Department of Physics and Astronomy, Iowa State University, Ames, IA 50011, USA}
\author{M.~K.~Daniel}\affiliation{Center for Astrophysics $|$ Harvard \& Smithsonian, Cambridge, MA 02138, USA}
\author{M.~Errando}\affiliation{Department of Physics, Washington University, St. Louis, MO 63130, USA}
\author{A.~Falcone}\affiliation{Department of Astronomy and Astrophysics, 525 Davey Lab, Pennsylvania State University, University Park, PA 16802, USA}
\author{K.~A.~Farrell}\affiliation{School of Physics, University College Dublin, Belfield, Dublin 4, Ireland}
\author{Q.~Feng}\affiliation{Department of Physics and Astronomy, Barnard College, Columbia University, NY 10027, USA}
\author{J.~P.~Finley}\affiliation{Department of Physics and Astronomy, Purdue University, West Lafayette, IN 47907, USA}
\author{L.~Fortson}\affiliation{School of Physics and Astronomy, University of Minnesota, Minneapolis, MN 55455, USA}
\author{A.~Furniss}\affiliation{Department of Physics, California State University - East Bay, Hayward, CA 94542, USA}
\author{A.~Gent}\affiliation{School of Physics and Center for Relativistic Astrophysics, Georgia Institute of Technology, 837 State Street NW, Atlanta, GA 30332-0430}
\author{C.~Giuri}\affiliation{DESY, Platanenallee 6, 15738 Zeuthen, Germany}
\author{D.~Hanna}\affiliation{Physics Department, McGill University, Montreal, QC H3A 2T8, Canada}
\author{T.~Hassan}\affiliation{DESY, Platanenallee 6, 15738 Zeuthen, Germany}
\author{O.~Hervet}\affiliation{Santa Cruz Institute for Particle Physics and Department of Physics, University of California, Santa Cruz, CA 95064, USA}
\author{J.~Holder}\affiliation{Department of Physics and Astronomy and the Bartol Research Institute, University of Delaware, Newark, DE 19716, USA}
\author{G.~Hughes}\affiliation{Center for Astrophysics $|$ Harvard \& Smithsonian, Cambridge, MA 02138, USA}
\author{T.~B.~Humensky}\affiliation{Physics Department, Columbia University, New York, NY 10027, USA}
\author{W.~Jin}\affiliation{Department of Physics and Astronomy, University of Alabama, Tuscaloosa, AL 35487, USA}
\author{P.~Kaaret}\affiliation{Department of Physics and Astronomy, University of Iowa, Van Allen Hall, Iowa City, IA 52242, USA}
\author{M.~Kertzman}\affiliation{Department of Physics and Astronomy, DePauw University, Greencastle, IN 46135-0037, USA}
\author{D.~Kieda}\affiliation{Department of Physics and Astronomy, University of Utah, Salt Lake City, UT 84112, USA}
\author{S.~Kumar}\affiliation{Physics Department, McGill University, Montreal, QC H3A 2T8, Canada}
\author{M.~J.~Lang}\affiliation{School of Physics, National University of Ireland Galway, University Road, Galway, Ireland}
\author{M.~Lundy}\affiliation{Physics Department, McGill University, Montreal, QC H3A 2T8, Canada}
\author{G.~Maier}\affiliation{DESY, Platanenallee 6, 15738 Zeuthen, Germany}
\author{C.~E~McGrath}\affiliation{School of Physics, University College Dublin, Belfield, Dublin 4, Ireland}
\author{P.~Moriarty}\affiliation{School of Physics, National University of Ireland Galway, University Road, Galway, Ireland}
\author{R.~Mukherjee}\affiliation{Department of Physics and Astronomy, Barnard College, Columbia University, NY 10027, USA}
\author{D.~Nieto}\affiliation{Institute of Particle and Cosmos Physics, Universidad Complutense de Madrid, 28040 Madrid, Spain}
\author{M.~Nievas-Rosillo}\affiliation{DESY, Platanenallee 6, 15738 Zeuthen, Germany}
\author{S.~O'Brien}\affiliation{Physics Department, McGill University, Montreal, QC H3A 2T8, Canada}
\author{R.~A.~Ong}\affiliation{Department of Physics and Astronomy, University of California, Los Angeles, CA 90095, USA}
\author{A.~N.~Otte}\affiliation{School of Physics and Center for Relativistic Astrophysics, Georgia Institute of Technology, 837 State Street NW, Atlanta, GA 30332-0430}
\author{N.~Park}\affiliation{WIPAC and Department of Physics, University of Wisconsin-Madison, Madison WI, USA}
\author{S.~Patel}\affiliation{Department of Physics and Astronomy, University of Iowa, Van Allen Hall, Iowa City, IA 52242, USA}
\author{K.~Pfrang}\affiliation{DESY, Platanenallee 6, 15738 Zeuthen, Germany}
\author{M.~Pohl}\affiliation{Institute of Physics and Astronomy, University of Potsdam, 14476 Potsdam-Golm, Germany and DESY, Platanenallee 6, 15738 Zeuthen, Germany}
\author{R.~R.~Prado}\affiliation{DESY, Platanenallee 6, 15738 Zeuthen, Germany}
\author{E.~Pueschel}\affiliation{DESY, Platanenallee 6, 15738 Zeuthen, Germany}
\author{J.~Quinn}\affiliation{School of Physics, University College Dublin, Belfield, Dublin 4, Ireland}
\author{K.~Ragan}\affiliation{Physics Department, McGill University, Montreal, QC H3A 2T8, Canada}
\author{P.~T.~Reynolds}\affiliation{Department of Physical Sciences, Munster Technological University, Bishopstown, Cork, T12 P928, Ireland}
\author{D.~Ribeiro}\affiliation{Physics Department, Columbia University, New York, NY 10027, USA}
\author{E.~Roache}\affiliation{Center for Astrophysics $|$ Harvard \& Smithsonian, Cambridge, MA 02138, USA}
\author{J.~L.~Ryan}\affiliation{Department of Physics and Astronomy, University of California, Los Angeles, CA 90095, USA}
\author{M.~Santander}\affiliation{Department of Physics and Astronomy, University of Alabama, Tuscaloosa, AL 35487, USA}
\author{G.~H.~Sembroski}\affiliation{Department of Physics and Astronomy, Purdue University, West Lafayette, IN 47907, USA}
\author{R.~Shang}\affiliation{Department of Physics and Astronomy, University of California, Los Angeles, CA 90095, USA}
\author{A.~Weinstein}\affiliation{Department of Physics and Astronomy, Iowa State University, Ames, IA 50011, USA}
\author{D.~A.~Williams}\affiliation{Santa Cruz Institute for Particle Physics and Department of Physics, University of California, Santa Cruz, CA 95064, USA}
\author{T.~J.~Williamson}\affiliation{Department of Physics and Astronomy and the Bartol Research Institute, University of Delaware, Newark, DE 19716, USA}

\collaboration{(VERITAS Collaboration)}

\author{I.~Bartos}
\affiliation{Department of Physics, University of Florida, Gainesville, FL 32611-8440, USA}

\collaboration{}%(University of Florida)}

\author{K.~R.~Corley}
\affiliation{Physics Department, Columbia University, New York, NY 10027, USA}

\author{S.~M\'arka}
\affiliation{Physics Department, Columbia University, New York, NY 10027, USA}

\author{Z.~M\'arka}
\affiliation{Columbia Astrophysics Laboratory, Columbia University in the City of New York, New York, NY 10027, USA}

\author{D.~Veske}
\affiliation{Physics Department, Columbia University, New York, NY 10027, USA}

\collaboration{(Columbia Experimental Gravity Group (GECo))}

\begin{abstract}
The recent discovery of electromagnetic signals in coincidence with neutron-star mergers has solidified the importance of multimessenger campaigns in studying the most energetic astrophysical events. Pioneering multimessenger observatories, such as LIGO/Virgo and IceCube, record many candidate signals below the detection significance threshold. These sub-threshold event candidates are promising targets for multimessenger studies, as the information provided by them may, when combined with contemporaneous gamma-ray observations, lead to significant detections. Here we describe a new method that uses such candidates to search for transient events using archival very-high-energy gamma-ray data from imaging atmospheric Cherenkov telescopes (IACTs). We demonstrate the application of this method to sub-threshold binary neutron star (BNS) merger candidates identified in Advanced LIGO’s first observing run. We identify eight hours of archival VERITAS observations coincident with seven BNS merger candidates and search them for TeV emission. No gamma-ray emission is detected; we calculate upper limits on the integral flux and compare them to a short gamma-ray burst model. We anticipate this search method to serve as a starting point for IACT searches with future LIGO/Virgo data releases as well as in other sub-threshold studies for multimessenger transients, such as IceCube neutrinos. Furthermore, it can be deployed immediately with other current-generation IACTs, and has the potential for real-time use that places minimal burden on experimental operations. Lastly, this method may serve as a pilot for studies with the Cherenkov Telescope Array, which has the potential to observe even larger fields of view in its divergent pointing mode.
\end{abstract}


\section{Introduction}
\label{sec:intro}

The recent association of electromagnetic counterparts to the first gravitational wave (GW) detection of a binary neutron star (BNS) merger by the Laser Interferometer Gravitational-Wave Observatory (LIGO)/Virgo and its partners worldwide has ushered in a new era of multimessenger astrophysics \citep{TheLIGOScientific:2017qsa}. This GW event, known as GW170817, was independently detected in gamma rays $\sim$ 1.7 s later and identified as a short gamma-ray burst \citep[GRB 170817A;][]{abbott2017multi}. GW170817 was followed-up extensively across the electromagnetic spectrum, and has enabled the identification of the host galaxy of the kilonova GW progenitor \citep{kasliwal2017illuminating, coulter2017swope}. Subsequent observations led to a great wealth of knowledge about kilonovae, nucleosynthesis, the origins of short gamma-ray bursts (GRBs), and more \citep{abbott2017multi}. 

In addition, IceCube's 2017 detection of a high-energy ($\sim$ 290 TeV) neutrino, IceCube-170922A, from the direction of the blazar TXS 0506+056 in a flaring state prompted a multiwavelength campaign to examine the possibility of blazars as a candidate source of high-energy neutrinos as well as very-high-energy (VHE;~>~100~GeV) gamma rays \citep{telescope2018multimessenger, icecube2018neutrino}. Initially, IceCube-170922A had only a 50\% probability of being astrophysical in nature, but reached the 3~$\sigma$ level when considered in the context of observations of the flaring gamma-ray state of TXS 0506+056 with the Large Area Telescope onboard the \textit{Fermi} satellite \citep[\textit{Fermi}-LAT;][]{atwood2009large}. Additional followup performed by the MAGIC, VERITAS, and H.E.S.S. Cherenkov telescope observatories led to the detection of a significant VHE gamma-ray signal from a direction consistent with the neutrino event by MAGIC \citep{telescope2018multimessenger} and VERITAS \citep{abeysekara2018veritas}. The follow up of such transient signals by ground-based imaging atmospheric Cherenkov telescopes (IACTs) thus holds great promise for the identification of VHE gamma-ray astrophysical counterparts.

While LIGO/Virgo and IceCube have made a number of high-significance detections independently, many of their potential signals fall short of the threshold for a detection, and thus do not trigger an alert \citep{LIGOScientific:2018mvr}. The plethora of candidates that are relegated to sub-threshold status presents the opportunity for re-examining archival data to look for correlated activity. The idea to take sub-threshold candidates from multimessenger observatories and perform real-time and archival coincidence searches therefore is a worthwhile approach, and one that continues to garner interest \citep{2001astro.ph.10349H,2003SPIE.4856..222M, 2008CQGra..25k4039A, 2008CQGra..25k4051A, 2010JPhCS.243a2001M, smith2013astrophysical, keivani2019ICRC, 2019arXiv190105486C, aartsen2020icecubesearch, 2020JCAP...05..016V, 2020ApJ...894..127W, solares2020astrophysical, 2020MNRAS.tmp.1969A}.

The catalogs of gravitational-wave events from the first (O1) through the first half of the third (O3a) observing run of Advanced LIGO and Virgo have been published \citep{LIGOScientific:2018mvr,2020gwtc2:arXiv201014527A}. The catalogs include 50 compact binary coalescences, of which two are classified as mergers of neutron stars. \citet{2020gwtc2popprop:arXiv201014533T} estimates a merger rate density of $23.9^{+14.9}_{-8.6}$~Gpc$^{-3}$ yr$^{-1}$ for binary black hole mergers, and of $320^{+490}_{-240}$~Gpc$^{-3}$ yr$^{-1}$ for binary neutron star mergers.

In the first observing run of Advanced LIGO alone, no BNS mergers were discovered; however, 103 sub-threshold BNS candidate events have been identified following the run's completion\footnote{\url{https://dcc.ligo.org/public/0158/P1900030/001/index.html}} \citep{magee2019sub}. The analysis of \citet{magee2019sub} assigns a false-alarm-rate (FAR) to potential BNS merger signals, and defines a sub-threshold candidate to be a signal with a FAR of less than one per day. Given their astrophysical probabilities, 1.63 of the 103 identified candidates are expected to be authentic -- in other words, GW signals from a BNS coalescence \citep{magee2019sub}. Despite the high noise and accidental contamination fraction, these candidates have the potential to be correlated with other multimessenger signals that could bolster the confidence in identifying real astrophysical events. In particular, a detection by an IACT, coincident in time and space, could help to increase the detection significance of a GW event. In the case a sub-threshold alert is released in real time, it is possible that such a near-real-time association by an IACT could trigger a campaign of further observations. 

VHE emission from long GRBs has been detected by IACTs \citep{magic2019teraelectronvolt, abdalla2019very, denaurois2019grb190829a}, further demonstrating the sensitivity of these ground-based gamma-ray observatories to short, high-energy transient events. Although GW detectors have all-sky sensitivity, the uncertainty in their localization of a source's point of origin remains large.\footnote{The median sky localization area (90\% credible region) is on the order of a few hundred square degrees for BNS mergers during O3 with the Advanced LIGO and Virgo (HLV) network \citep{abbott2018prospects}.} On the other hand, IACTs have a small field of view (FoV) but comparatively excellent localization.\footnote{The angular resolution of VERITAS is of order~0.13$\degree$ at 200~GeV, with a source location accuracy of 50~arcseconds for a sufficiently bright detection. Details on VERITAS performance can be found at \url{https://veritas.sao.arizona.edu/about-veritas/veritas-specifications}.}  The association of VHE gamma rays with GW events would have interesting astrophysical implications -- in particular, source localization and the identification of the production mechanism of VHE photons. The full multimessenger view is important to understand the nature of the source and environment of these events, and the detection of their electromagnetic counterparts will provide valuable information on the characteristics of the central engine and perhaps even the origin of short GRBs.

In this work, we describe and demonstrate a transient archival search method using IACT data from VERITAS to search for serendipitous coincidences with sub-threshold BNS candidates from Advanced LIGO's first observing run. If VERITAS is observing within a candidate’s high localization probability region at the time of its measurement with LIGO and sees flaring activity, this could potentially improve confidence in a LIGO detection and provide improved localization. The primary goal of this work is to establish the methodology, capability, and potential for discovery of this method. In addition, this study could possibly lead to follow-up of new sub-threshold triggers in future observing runs in real time, not just with VERITAS, but also with other current generation IACTs, H.E.S.S. and MAGIC \citep{aharonian2006HESS,aleksic2016major}.  This work may also serve as a prototype for future studies with CTA (the Cherenkov Telescope Array), as the improvements in sensitivity of GW detectors coupled with CTA's wider FoV and better sensitivity will increase the likelihood of having serendipitous coverage of the core, or highest probability region, of a GW candidate event over time. 

\section{GW Observations and VERITAS Search Method} \label{sec:method}
LIGO comprises two kilometer-scale gravitational-wave detectors in Hanford, Washington, USA and Livingston, Louisiana, USA \citep{abbott2009ligo}. The inaugural observing run (O1) of the Advanced LIGO \citep{aasi2015advanced} detectors at these sites, spanning from September 12, 2015 to January 19, 2016, yielded the first ever detection of a binary black hole merger, but did not identify any unambiguous gravitational wave signals associated with the merger of binary neutron stars. However, 103 sub-threshold BNS merger candidates with a false-alarm-rate of less than one per day have been identified \citep{magee2019sub}.

To identify any coincident VERITAS observations within these candidates, we required that a point within the most credible region at 90\% confidence (i.e. the 90\% localization region) of the LIGO BNS merger candidate also be within the FoV of a VERITAS observation pointing (at nominal operating voltage and zenith angle of less than 55$\degree$) that overlapped with the time window $-10 \leq t_{0} \leq 10^4$~seconds, where $t_{0}$ is the coalescence time of the GW event candidate. This window was chosen based on the suggestion that VERITAS, if aligned with the jet axis of a short-duration GRB, would be sensitive to such an event with VHE plateau emission extending to $10^4$~seconds \citep{murase2018double}. A multi-timescale search is an interesting prospect; however, it is not one we consider in the scope of this paper. We utilize the credible region at 90\% to strike a balance between the diminishing returns of an expanding credible region and the VERITAS analysis efficiency. Out of the 103 sub-threshold BNS candidates, 7 were found to have temporal and spatial coincidence with 11 sets of archival VERITAS observations. Some candidates had multiple VERITAS observations found to be in spatial and temporal coincidence, as documented in \autoref{tab:results} and \hyperref[app:ligo]{Appendix} (see \autoref{fig:C3} and \autoref{fig:C5}). These candidates and their coverage in $\sim$~8 hours of VERITAS target observations are documented in \autoref{tab:ligo_candidates} and shown in the \hyperref[app:ligo]{Appendix}. From these, the runlist of serendipitous spatially and temporally coincident observations performed by VERITAS was generated.

\begin{table*}[h]
\centering
\caption{The 7 sub-threshold BNS merger candidates found to have spatial and temporal coincidence with at least one VERITAS observation. For these candidates, we present the candidate label, LIGO event ID ($t_0$ in UTC), false-alarm rate (FAR), signal-to-noise ratio (S/N),  probability the candidate is astrophysical in origin, and area of the 90\% confidence region \citep{magee2019sub}. We also provide the time of the first spatially coincident VERITAS observation within the temporal window with respect to the candidate's $t_0$ ($t_{first}$), the total spatially coincident VERITAS exposure time within the temporal window ($t_{coinc}$), and the probability the counterpart, if authentic, fell within the VERITAS FoV during those observations. Candidates marked by H and L were single-detector triggers from LIGO-Hanford and LIGO-Livingston respectively.} 
\begin{tabular}{llllllrrr}
\hline
\multirow{3}{*}{\shortstack[l]{Candidate \\Label}} & \multirow{3}{*}{\shortstack[l]{LIGO BNS Candidate\\ Event ID}} & \multicolumn{4}{c}{LIGO} & \multicolumn{3}{c}{VERITAS} \\ \cmidrule(lr){3-6}\cmidrule(lr){7-9}
&& \shortstack[l]{FAR \\ (yr$^{-1}$)}  & S/N   & \shortstack[l]{p-astro \\ ($10^{-3}$)} & \shortstack[l]{Area \\(deg$^2$)}  & $t_{first}$ & $t_{coinc}$ & \shortstack[l]{Coverage \\Probability} \\
\hline
C1      & 2015Oct12T02:40:22.39 & 142.27    & 8.42  & 3.82   & 2321  & -0:11:17 &   0:18:53 & 0.22\%    \\
C2$^L$  & 2015Oct24T09:03:52.00 & 7.52      & 9.69  & 79.6   & 24218 & 1:33:08  &   1:11:08 & 0.06\%    \\
C3$^H$  & 2015Nov17T06:34:02.07 & 7.52      & 8.84  & 181    & 24221 & -0:08:02 &   2:37:43 & 0.18\%    \\
C4      & 2015Dec04T01:53:39.14 & 225.02    & 9.09  & 2.5    & 2909  & 0:16:20  &   1:00:00 & 0.19\%    \\
C5$^L$  & 2015Dec06T06:50:38.17 & 77.45     & 7.72  & 6.64   & 24264 & -0:09:02 &   2:10:18 & 0.15\%    \\
C6      & 2015Dec09T07:25:24.68 & 141.65    & 7.85  & 3.84   & 2606  & 1:36:25  &   0:15:00 & 0.03\%    \\
C7      & 2016Jan02T02:47:29.35 & 356.13    & 7.51  & 1.63   & 3487  & 1:44:55  &   0:30:00 & 0.18\%    \\
\hline
\end{tabular}
\label{tab:ligo_candidates}
\end{table*}

\begin{figure}[t!]
    \includegraphics[width=\columnwidth]{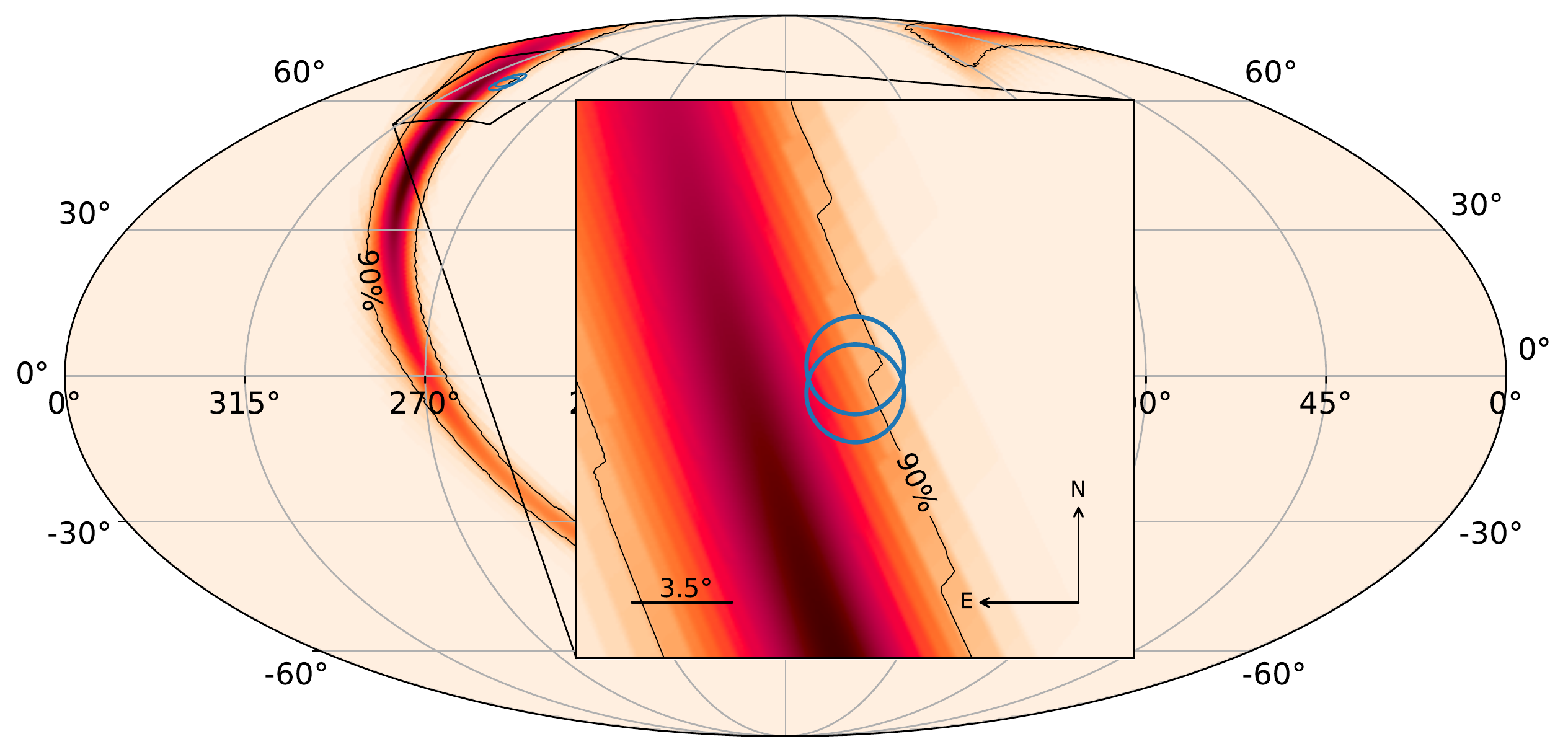}
    \caption{The localization probability map, with 90\% localization contours shown in black, for sub-threshold LIGO BNS candidate C1 (from \autoref{tab:ligo_candidates}) presented in equatorial coordinates with two VERITAS observations that overlap spatially and temporally overlaid (blue circles).}
    \label{fig:1ES_example}
\end{figure}

The result of this algorithm in identifying coincident observations is demonstrated in \autoref{fig:1ES_example}. The VERITAS observations in the figure were North/South wobbled (0.5$\degree$ offset from the target) observations of the blazar 1ES 1959+650 and each lasted 15 minutes. The LIGO sub-threshold candidate $t_0$ was 02:40:22 UTC. The first VERITAS observation, a North wobble, was performed from 02:29:05 until 02:44:05 UTC and had a probability of encompassing the LIGO candidate of 0.13\%. The second, a South wobble, was performed from 03:17:07 until 03:32:07 UTC and had a probability of encompassing the LIGO candidate of 0.17\%. Here, we see a set of observations that not only passes our temporal and spatial cuts, but also includes data contemporaneous with the $t_{0}$ of the candidate: a fortuity made possible by our use of archival data.

\begin{figure}[h]
    \centering
    \includegraphics[width=\columnwidth]{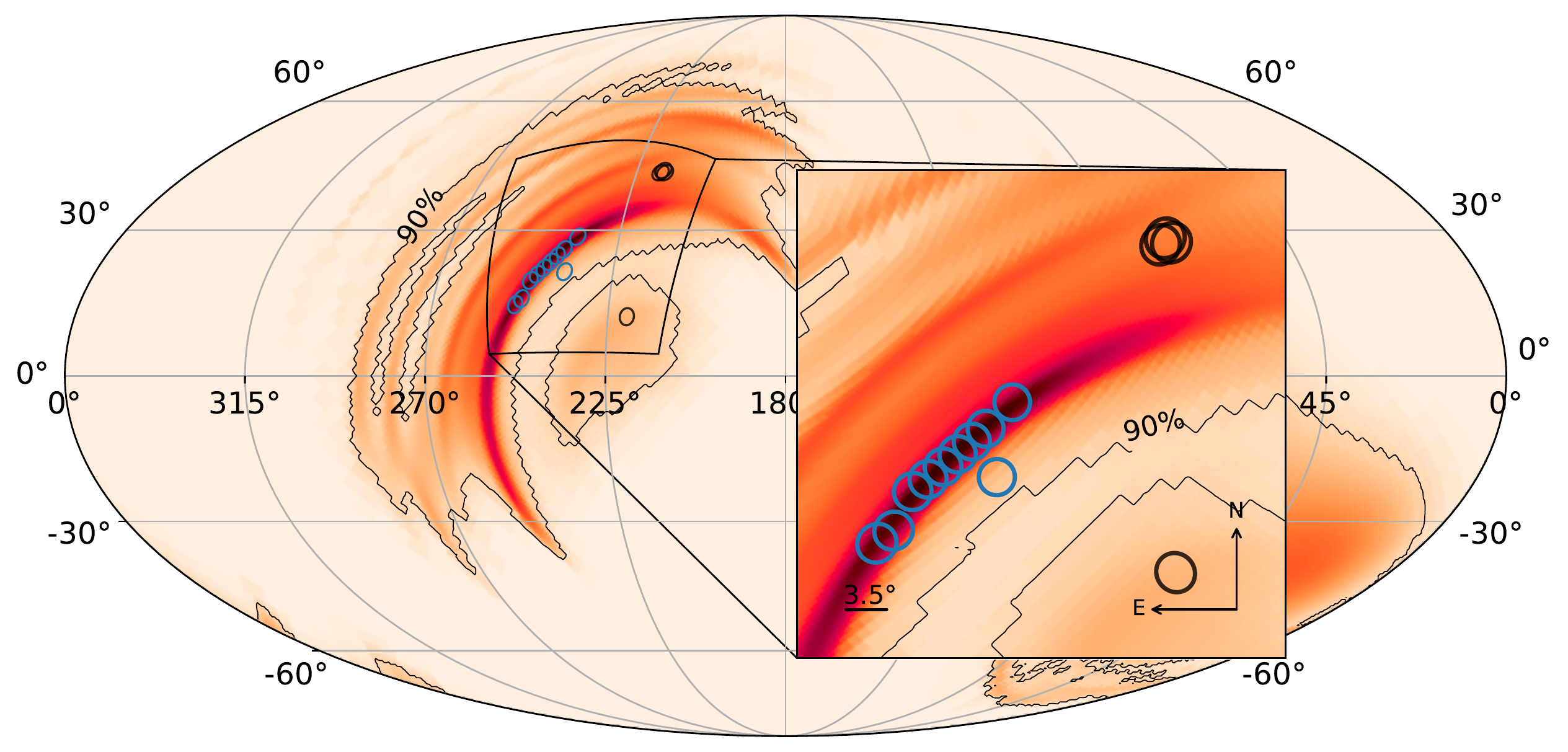}
    \caption{The localization probability map, with 90\% localization contours shown in black, for 2019 BNS merger event G330561 presented in equatorial coordinates. Ten VERITAS observations (blue circles) trace the highest localization probability region of the LIGO skymap. Four serendipitous VERITAS observations (black circles) also passed the spatial and temporal cuts of the algorithm and were taken prior to initial VERITAS follow-up.}
    \label{fig:G330561}
\end{figure}

This algorithm was cross-checked using the intentional VERITAS follow up of the G330561 BNS merger event in 2019 \citep{gcn_g330561}, shown in \autoref{fig:G330561}. In addition to the 10 triggered pointings chosen to observe at the points of highest LIGO localization probability, the coincidence algorithm also identified 4 additional spatially-coincident serendipitous pointings taken within the time window defined above.

\section{VERITAS Analysis}
\label{sec:analysis}
VERITAS \citep{holder2006first} is a VHE gamma-ray telescope array located at the Fred Lawrence Whipple Observatory in southern Arizona, USA. It consists of four imaging atmospheric Cherenkov telescopes which use tesselated, 12-meter diameter reflectors to collect Cherenkov light created by particle cascades, or air showers, initiated by gamma rays and cosmic rays in the Earth's atmosphere. Each telescope records images of these showers using a focal-plane camera covering a 3.5$\degree$ FoV. These images are then used to reconstruct the direction and energy of the initiating gamma rays. The array is sensitive to gamma rays in the 80~GeV to 30~TeV energy range and is able to make a 5$\sigma$ detection of the Crab Nebula in under a minute.\footnote{See \url{https://veritas.sao.arizona.edu/about-veritas/veritas-specifications}.} VERITAS has long had a program to follow up on transient events from multimessenger observatories such as IceCube and LIGO \citep{aartsen2017multiwavelength,telescope2018multimessenger, santander2019recent}.

For each set of VERITAS observations passing the search algorithm described in the previous section, we perform a point-source analysis using the standard analysis pipeline for VERITAS data \citep{acciari2008veritas} to search for a significant excess. After the data has been calibrated and cleaned \citep{daniel2008veritas, cogan2006thesis}, air shower images are parameterized using the Hillas moment analysis \citep{hillas1985cerenkov}. The scaled parameters, used for event selection, are calculated from the moment analysis \citep{aharonian1997potential, krawczynski2006gamma}. Event selection cuts are chosen \textit{a priori}. Given the uncertainty of the spectral features of short-duration GRBs at VHE and the extragalactic nature of such events, we opt in this analysis for ``soft cuts,'' which have the best sensitivity at lower energies, around a few hundred GeV, and are optimized for sources with a soft spectrum (spectral index of -3.5 to -4.0) \citep{park2015performance}. To produce a significance map of the observations, the background is estimated using a ring background model \citep[RBM;][]{berge2007background} with a point source search integration radius of $\theta_{int}=0.17\degree$. The significance of the deviation from the RBM background is given by Equation 17 of \citet{lima1983analysis}. For each set of observations, this significance is calculated at each point in a grid spaced at 0.025$\degree$ to generate a skymap of observed significances. For this analysis, each VERITAS skymap is searched for significant excesses (excluding points $< 0.5\degree$ from known sources, $ < 0.3\degree$ from bright stars, and $< 0.2\degree$ from the edge of the camera to remove effects due to high fluctuations). Given the large uncertanties of the BNS candidate localization skymaps relative to the VERITAS FoV, we also generate a skymap of the bounded upper limits on the integral flux (at 99\% confidence) using the method of \citet{rolke2005limits}, assuming a soft spectral index of -3.5 \citep{abdalla2019very}. For each skymap, we report the geometric mean of the upper limits on the integral flux. Confirmation of results was obtained using an independent secondary analysis, as described in \citet{maier2017evndisp}.

\section{Results} \label{sec:results}
\begin{figure*}[!hbt]
    \centering
    \subfloat[\label{fig:IFULmap}]{%
        \includegraphics[height=7.5cm]{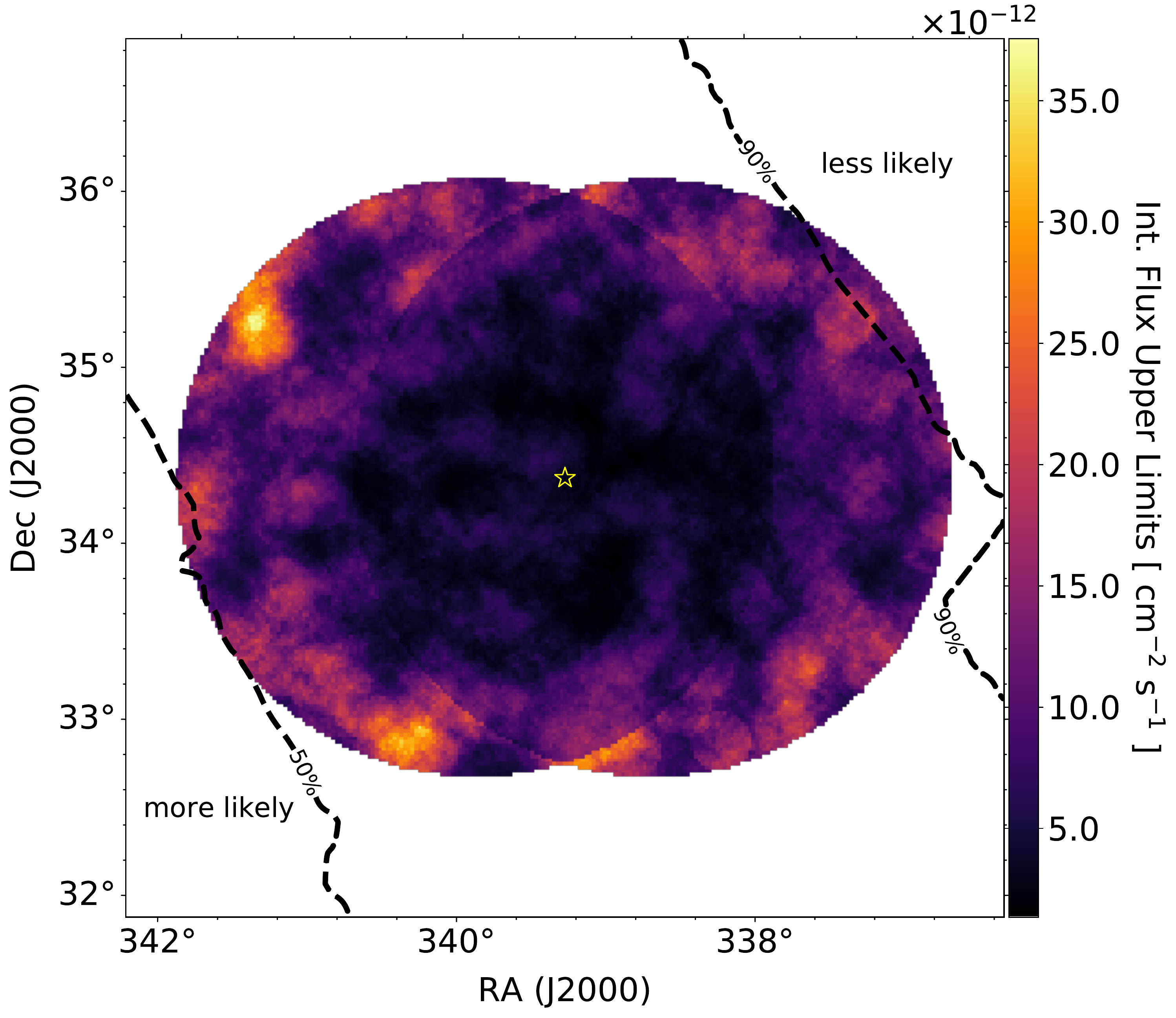}%
    }
    \hfill
    \subfloat[\label{fig:IFULhist}]{%
        \includegraphics[height=7.5cm]{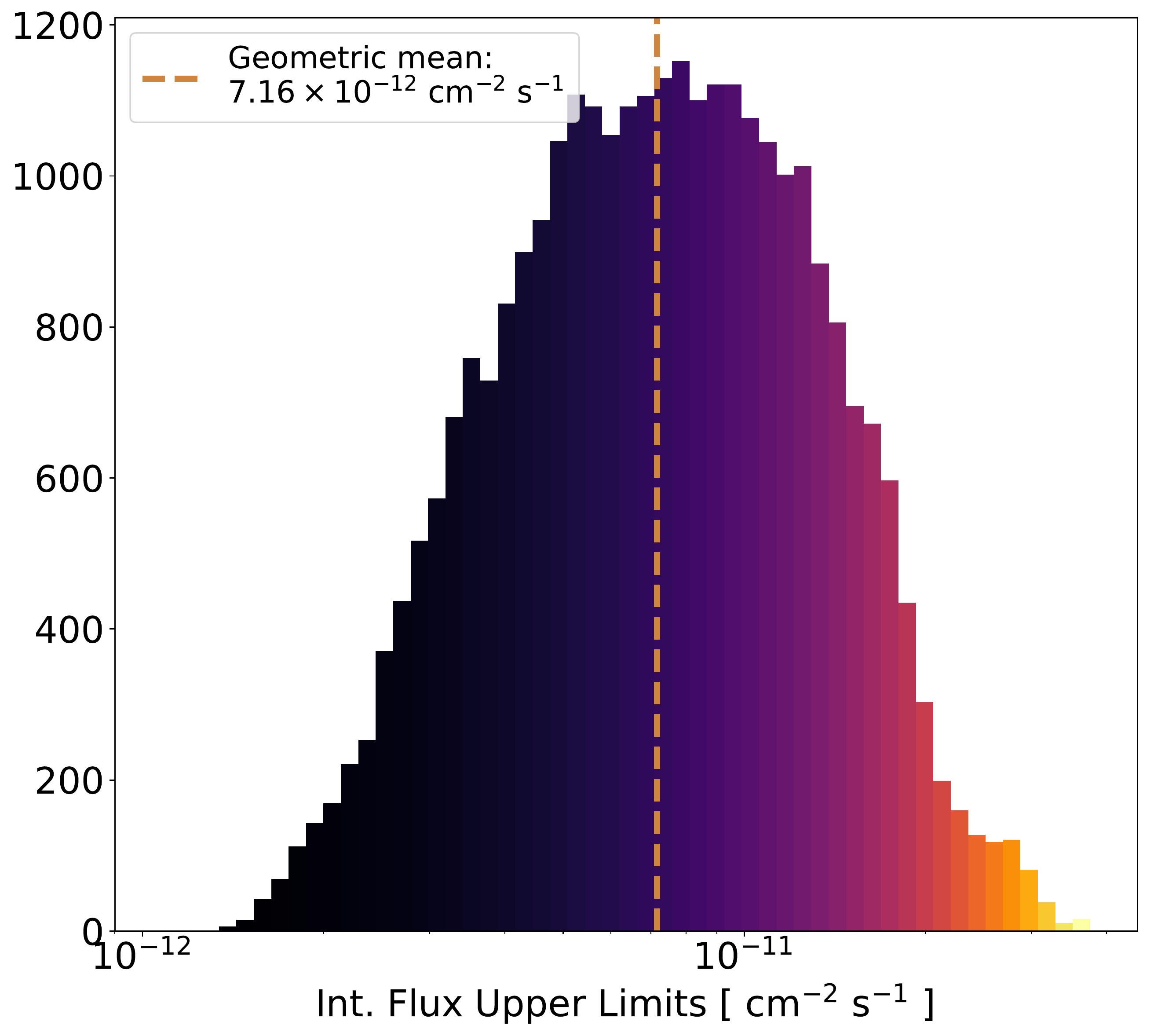}%
    }
    \caption{Results from VERITAS observations of the target SN~2014c (marked with a star), made in coincidence with the LIGO BNS candidate C4. Panel~\subref{fig:IFULmap} shows a skymap of upper limits on the integral flux over [0.24~TeV,~30~TeV] for the two pointings in the observation, with the 50\% and 90\% localization contours of the GW candidate overlaid. The distribution of the upper limits from \subref{fig:IFULmap} is shown in \subref{fig:IFULhist}; both use the same colormap to better connect the data visually. The geometric mean of this distribution is indicated by the vertical line.}
\end{figure*}

Under the search conditions described in \autoref{sec:analysis}, our results reveal no significant excesses with a pre-trial significance above 5~$\sigma$. Furthermore, the $\sim98$ square degree sky region scanned across the 11 archival sets of VERITAS observations subjects our search to a large number of trials, necessarily imposing a penalty on any pre-trial significances observed. Two independent estimations of the incurred trials indicate that all post-trial significances observed are fully compatible with the background hypothesis \citep{funk2005thesis}, and thus, are consistent with the distribution of significances expected with no signal present \citep{abeysekara2018very}. We subsequently generate skymaps of the bounded upper limits on the integral flux at 99\% confidence and report their geometric means, defined as the $n$th root of a product of $n$ numbers, in \autoref{tab:results}. They may be compared to those shown in \citet{abdalla2017tev, ashka2019searches, seglar2019searches}, with an important distinction that our observations are motivated by sub-threshold candidates instead of reliably identified merger events. The geometric mean is extracted from the distribution of upper limits in each skymap, where points near known TeV sources, bright stars, or near the edge of the FoV have been excluded. The most constraining upper limit in each analysis is typically 20\% of the geometric mean. In \autoref{fig:IFULmap}, we provide an example integral flux upper limit skymap for the VERITAS observations of SN 2014c that took place coincidentally with BNS merger candidate C4. We additionally show the distribution and geometric mean of the integral flux upper limits in the skymap in \autoref{fig:IFULhist}, with sources and bright stars excluded from the distribution.

\begin{table*}[h]
\centering
\caption{VERITAS results for the geometric mean of the 99\% bounded \citet{rolke2005limits} upper limits on the integral flux over E$_{min}$ to E$_{max}$=30 TeV within each VERITAS observation skymap. We provide the candidate label (superscripts as defined in \autoref{tab:ligo_candidates}), the target of the archival VERITAS observation, the livetime spent on each target, and the geometric mean of the integral flux upper limits in that skymap. Two values of the upper limit are provided, one "common" value where all E$_{min}$ are set to 240 GeV, the maximum threshold energy (E$_{thresh}$) of all VERITAS observations, and one "unique" value where the E$_{min}$ for each BNS candidate is set to the max VERITAS E$_{thresh}$ of that candidate. Redundant cases have been omitted with a dash.}
\begin{tabular}{lllccc}
\hline
\noalign{\vskip 5pt}  
\multirow{3}{*}{\shortstack[l]{Candidate\\Label}} & \multirow{3}{*}{VERITAS target} & \multirow{3}{*}{\shortstack[l]{Observation\\Livetime}}  & Common E$_{min}$=240 GeV & \multicolumn{2}{c}{Unique E$_{min}$} \\ \cmidrule(lr){4-4} \cmidrule(lr){5-6}
&&& \shortstack[c]{Int. Flux UL \\ \relax (10$^{-12}$ cm$^{-2}$ s$^{-1}$)} & E$_{min}$ (GeV) & \shortstack[c]{Int. Flux UL\\ \relax (10$^{-12}$ cm$^{-2}$s$^{-1}$) } \\
\hline
C1      & 1ES 1959+650          & 0:30:00  & 21   & 240   & -         \\ 
\arrayrulecolor{black!30}\specialrule{.001em}{0em}{0em} 
C2$^L$  & PSR B0355+54 tail     & 1:29:08  & 12   & 220   & 16      \\
\specialrule{.001em}{0em}{0em} 
C3$^H$  & 3C 66A                & 1:00:00  & 8.2   & 240   & -         \\ 
C3$^H$  & 2FHL J0245.6+6605     & 1:00:01  & 17   & 240   & -         \\ 
C3$^H$  & Crab Nebula           & 1:00:01  & 5.8   & 240   & -         \\ 
\specialrule{.001em}{0em}{0em} 
C4      & SN 2014c              & 0:59:09  & 7.2   & 140   & 29       \\
\specialrule{.001em}{0em}{0em} 
C5$^L$  & 2FHL J0431.2+5553e    & 1:00:01  & 12   & 170   & 30      \\ 
C5$^L$  & 3C 66A                & 0:20:01  & 18   & 170   & 45       \\ 
C5$^L$  & Crab Nebula           & 0:59:08  & 4.9   & 170   & 12      \\ 
\specialrule{.001em}{0em}{0em} 
C6      & 1ES 0806+524          & 0:15:00  & 14   & 150   & 46      \\ 
\specialrule{.001em}{0em}{0em}\arrayrulecolor{black}
C7      & VER J0521+211         & 0:30:00  & 6.5   & 140   & 26      \\  \hline
\end{tabular}

\label{tab:results}
\end{table*}

\section{Discussion and Future Perspectives} \label{sec:discussion}
From the VERITAS upper limits on the integral flux, we can calculate upper limits on the fluence for each of these observations by converting to an integral energy flux over the range [0.24 TeV, 30 TeV] (the "common" energy interval discussed in \autoref{tab:results}) and multiplying by the observation livetime with the assumption of constant flux. We can then compare these values to the predicted time evolution of a short GRB using the model in \citet{Bartos:2014thv, bartos2019gravitational} for the afterglow of a representative burst, GRB~090510. This model extrapolates GRB~090510's observed emission at $\sim$~100~MeV and $\sim$~100~s by \textit{Fermi}-LAT out to longer durations and higher energies assuming a simple synchrotron component emerging from an electron population accelerated by the external forward shock \citep{Bartos:2014thv, kumar2015physics, corsi2010gev}. For this comparison, we model the GRB at a distance of 75 Mpc, the BNS merger range \citep{chen2017distance} of the Advanced LIGO detectors during the O1 observing run \citep{martynov2016sensitivity}, and adapt the model to cover the same [0.24~TeV,~30~TeV] energy interval as above. The model is then adjusted to include a quadratic fast rise from $t_0$~to~$t_{peak}$ such that $t_{peak}-t_0~=~\textrm{2.19 s}$ \citep{ghirlanda2010onset}. We choose this number to account for the delay in the peak of GeV emission from the neutron star coalescence, motivated by the 1.74~s delay between $t_0$ of GW170817 and the GBM trigger for GRB 170817A \citep{abbott2017gravitational} and the 0.82~s delay between the GBM trigger and the peak of GeV emission for GRB~090510  \citep{ghirlanda2010onset}. These $\Delta t$s are then redshift-corrected to the model's placement at 75 Mpc, yielding 1.75~s and 0.44~s respectively.

It should be noted that the physics of generating photons of the order 100 GeV may differ from that of generating high-MeV/low-GeV photons. In particular, if this simple synchrotron model is assumed, it requires exceptionally high bulk Lorentz factors ($\Gamma$~\textgreater~1500) to produce photons at the multi-GeV/TeV level \citep{inoue2013gamma, Bartos:2014thv, ajello2020fermi}, likely suggesting a cutoff at these energies. However, as this cutoff remains uncertain \citep{wood2016all}, it is worthwhile to consider the prospects in the speculative case where this emission continues out to TeV energies. It is also worth mentioning that the inverse Compton process dominates TeV emission observed in long GRBs \citep{veres2019observation}. As it is not clear how comparable the emission processes of long and short GRBs are, this places an additional caveat on the model considered in this paper.

\begin{figure*}[t!]
    \centering
    \includegraphics[width=0.7\textwidth]{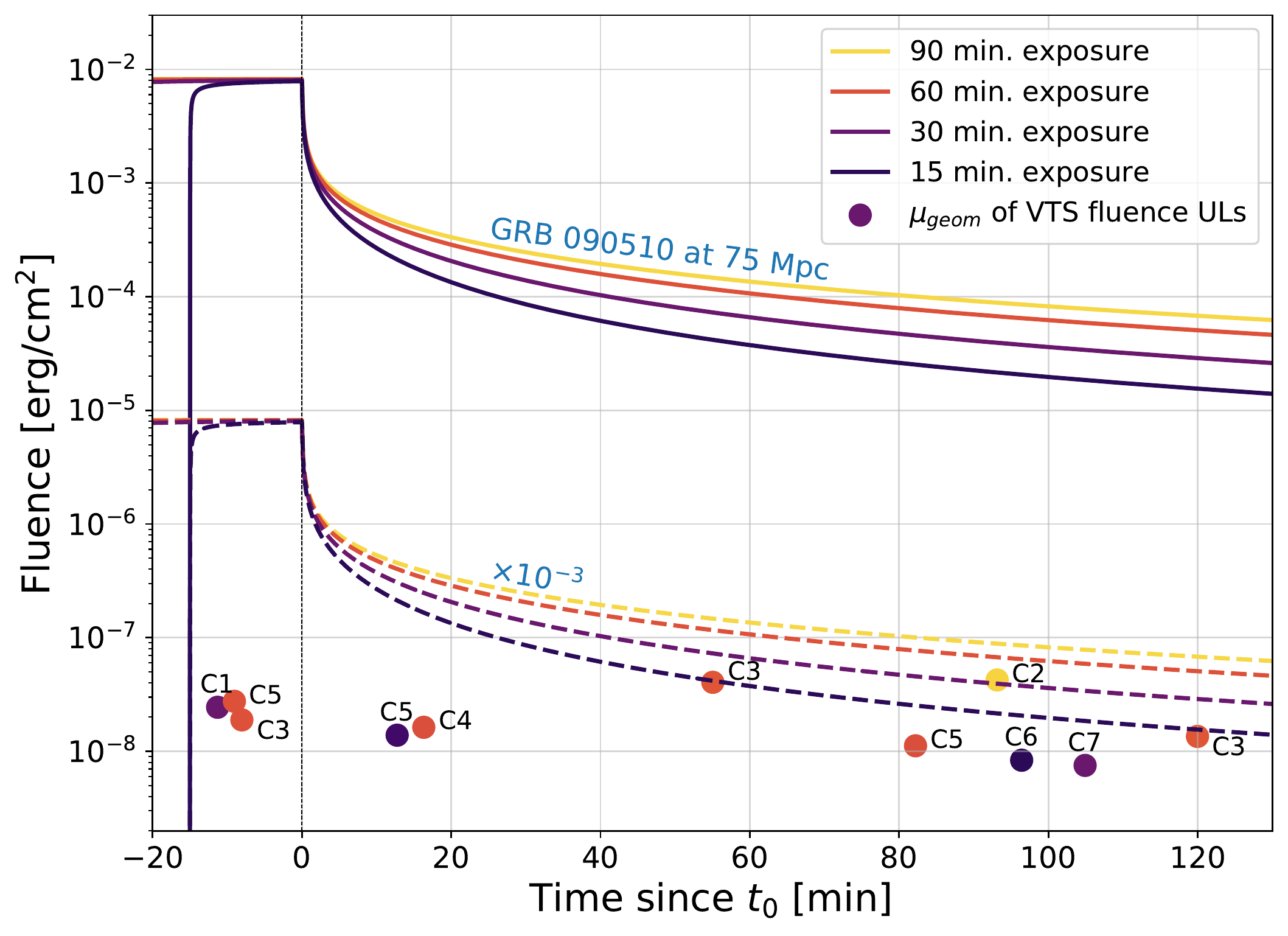}
    \caption{The curves in this plot depict the estimated fluence evolution of the \citet{Bartos:2014thv, bartos2019gravitational} GRB model placed at 75 Mpc over the [0.24~TeV,~30~TeV] energy band. Each point of the curve is the energy flux integrated from that point in time to $n$ minutes later, where $n$ is determined by the legend. Also shown with dashed lines are fluence estimates from a source 3 orders of magnitude weaker. The geometric means ($\mu_{geom}$) of the VERITAS upper limits on the fluence in each skymap (calculated from the common $E_{min}$ upper limits on the integral flux shown in \autoref{tab:results}) are plotted at the start time of each VERITAS observation with respect to $t_0$. Each observation is tagged with the label of its corresponding BNS candidate (see \autoref{tab:ligo_candidates}). The colors of the $\mu_{geom}$ points are coded by their observation livetime on the same color scale as the model curve for easy comparison.}
    \label{fig:fluenceULs}
\end{figure*}

We show the results of the model comparison in \autoref{fig:fluenceULs}. For each VERITAS skymap, we plot the geometric mean of the bounded \citet{rolke2005limits} 99\% upper limits on the livetime fluence from the observed region of sky. All VERITAS upper limits fall orders of magnitude below the estimated fluence from the GRB~090510 placed at a distance of 75 Mpc. Consequently, we constrain any VHE counterparts in the VERITAS FoV at the time of observation with emission similar to that predicted for GRB~090510 by the model. As mentioned in \autoref{sec:intro}, it is unlikely that any of these overlapping observations of BNS candidates in fact contain real BNS mergers. Therefore, the most plausible explanation is that no GRBs associated with BNS merger events were observed. If, however, a BNS merger were in the FoV for any of these observations, then these results suggest that its resulting GRB be far less luminous than predicted for GRB~090510 using this model. This can be due either to orientation (jet not aligned with the line of sight), a cutoff below 240 GeV, or potentially variations in the jet properties \citep{zhang_2018_grbphysics}.

The capability to carry out transient archival searches is compelling as it provides a potential route to the observation of the onset of VHE emission, if it exists, for GRBs associated with BNS mergers. In particular, such an observation has great utility in placing constraints on the various models \citep{zhang_2018_grbphysics} for the structured emission in the jet from such an event. Even without a detection, this method may provide the opportunity to constrain VHE emission models of short GRBs with a significant build-up of candidates in future data releases.

Using the probability of astrophysical origin from \autoref{tab:ligo_candidates}, we infer the probability that at least one of the merger candidates tagged by the coincidence algorithm was a real BNS merger event to be 26\%. We further estimate, using the VERITAS coverage of each candidate, that the probability that at least one truly astrophysical merger was observed by VERITAS with exact spatial coincidence in the search time window to be 0.04\%. Despite the low probability of authentic GW signals in these O1 candidates, and the limited GW localization probability area covered by their coincident VERITAS observations, it is important to bear in mind that future data comes at no additional cost and minimal to no burden on observing programs. Additionally, the situation will only improve with the onset of new and upgraded instruments, and just one positive result would be very high impact. In the second and third observing runs from Advanced LIGO, we will continue to identify new sub-threshold BNS merger candidates all while probing greater distances with enhanced localization. If we scale our O1 performance to O2 and O3, accounting for instrumental improvements and changing duty cycles, we predict the accumulation of $\sim$ 70 or more additional LIGO sub-threshold BNS merger candidates that VERITAS will coincidentally observe some fraction of as a part of its routine observation, a factor of 10 improvement on our current number. It is straightforward to generalize this method to the other currently operating IACTs, H.E.S.S. and MAGIC, and with their inclusion we could also expect an additional 2-to-3 fold increase. While these improvements alone may not be enough to place meaningful constraints, even one particularly serendipitous observation covering a large GW localization probability area of a sub-threshold candidate could significantly bridge the gap necessary to accomplish this.

Beyond the advancements in LIGO's sensitivity, this paper's method will also benefit from the imminent era of astronomy with the Cherenkov Telescope Array. CTA's sky coverage will expand on and complement the coverage provided by the current generation of IACTs: VERITAS, MAGIC, and H.E.S.S. \citep{2019scta.book.....C,bartos2019gravitational}. Additionally, standard CTA operations will have an anticipated FoV of $\sim 8\degree$, a factor of $\sim$ 5 improvement in area over the $3.5 \degree$ FoV provided by VERITAS. Note, however, that the Large-Sized Telescopes (LSTs), sensitive to the 20 GeV - 3 TeV energy range, will have a FoV of only $4.5 \degree$ \citep{barrio2020status}. This paper's method will also have enhanced utility in the context of CTA's proposed divergent pointing mode for extragalactic surveys. A divergent pointing mode, without LSTs, will increase CTA's already expansive FoV to $\sim 14 \degree$ or larger \citep{gerard2016divergent}, a factor of 16 improvement in sky coverage over VERITAS, at the cost of angular and energy resolution and average instantaneous sensitivity \citep{donini2019cherenkov}.

While the work in this paper was done with archival data, it could also be done in near real-time, with minimal burden on the operations of currently operating IACTs. In particular, with communication of sub-threshold BNS candidates to Cherenkov telescopes, IACTs could slightly adjust observing schedules to prioritize the already planned observations which encompass the regions of higher localization probability. Such adjustments can be made according to a sliding scale that determines how much modification of the schedule is warranted. Observations of this kind would at most require trivial modification to existing real-time analysis tools. With these dedicated analyses, if an interesting gamma-ray excess were to be identified overlapping with the time and localization of a sub-threshold BNS merger candidate alert, IACTs could send out subsequent alerts for further follow-up.

\section{Conclusion} \label{sec:conclusion}
In this paper, we pioneered a novel method with VERITAS to study sub-threshold BNS merger candidates from Advanced LIGO's first observing run. Out of 103 candidates, we identified 7 with VERITAS observations coincident spatially and temporally. With these observations, we carried out a search for TeV emission and, finding none, provided characteristic upper limits on the integral flux for the coincident regions. From these upper limits, we calculated the fluence and compared these to a short GRB afterglow model extrapolated from GRB~090510 to higher energies and longer durations, and placed at the O1 BNS range. We report that all characteristic upper limits from VERITAS fall orders of magnitude below this model, thus suggesting that, if synchrotron emission from the forward shock extends up to the VHE range, we did not observe such an event, or that we observed a much less luminous one under the assumption that the event was, in fact, of astrophysical origin and it was located within the VERITAS FoV.

Our method demonstrates the prospects for the use of archival data to investigate coincident observations of BNS merger candidates and to build-up constraining observations of the VHE emission of such events over time. Additionally, the strategy developed here is interesting not only for sub-threshold candidates, but also for future high-confidence, single-detector GW events where the localization is poor, thus prohibiting targeted IACT follow-up. Future studies, especially with near real-time collaboration between experiments, may prove a boon for the field of multimessenger astronomy, given their potential to assist in upgrading sub-threshold candidates to detections. It will benefit such studies to build on this method with the inclusion of joint calculations to estimate significance and upper limits. A general framework for this addition may be found in \citet{veske2020mmsearch}. 

\acknowledgments
{This research is supported by grants from the U.S. Department of Energy Office of Science, the U.S. National Science Foundation and the Smithsonian Institution, by NSERC in Canada, and by the Helmholtz Association in Germany. This research used resources provided by the Open Science Grid, which is supported by the National Science Foundation and the U.S. Department of Energy's Office of Science, and resources of the National Energy Research Scientific Computing Center (NERSC), a U.S. Department of Energy Office of Science User Facility operated under Contract No. DE-AC02-05CH11231. We acknowledge the excellent work of the technical support staff at the Fred Lawrence Whipple Observatory and at the collaborating institutions in the construction and operation of the instrument.

The Columbia and Barnard authors thank Columbia University in the City of New York and Barnard College for their support, and acknowledge the generous support of the National Science Foundation under grant PHY-1806554.
Their Alabama colleagues thank the University of Alabama and acknowledge the support of the National Science Foundation under grant PHY-1914579.
I.~Bartos acknowledges the support of the National Science Foundation under grant PHY-1911796, the Alfred P. Sloan Foundation, and the University of Florida.
The Columbia Experimental Gravity group is grateful for the generous support of the National Science Foundation under grant PHY-1708028.}

\bibliography{references,biball}

\phantomsection
\appendix
\label{app:ligo}
\begin{figure*}[h!]
    \centering
    \includegraphics[width=\textwidth]{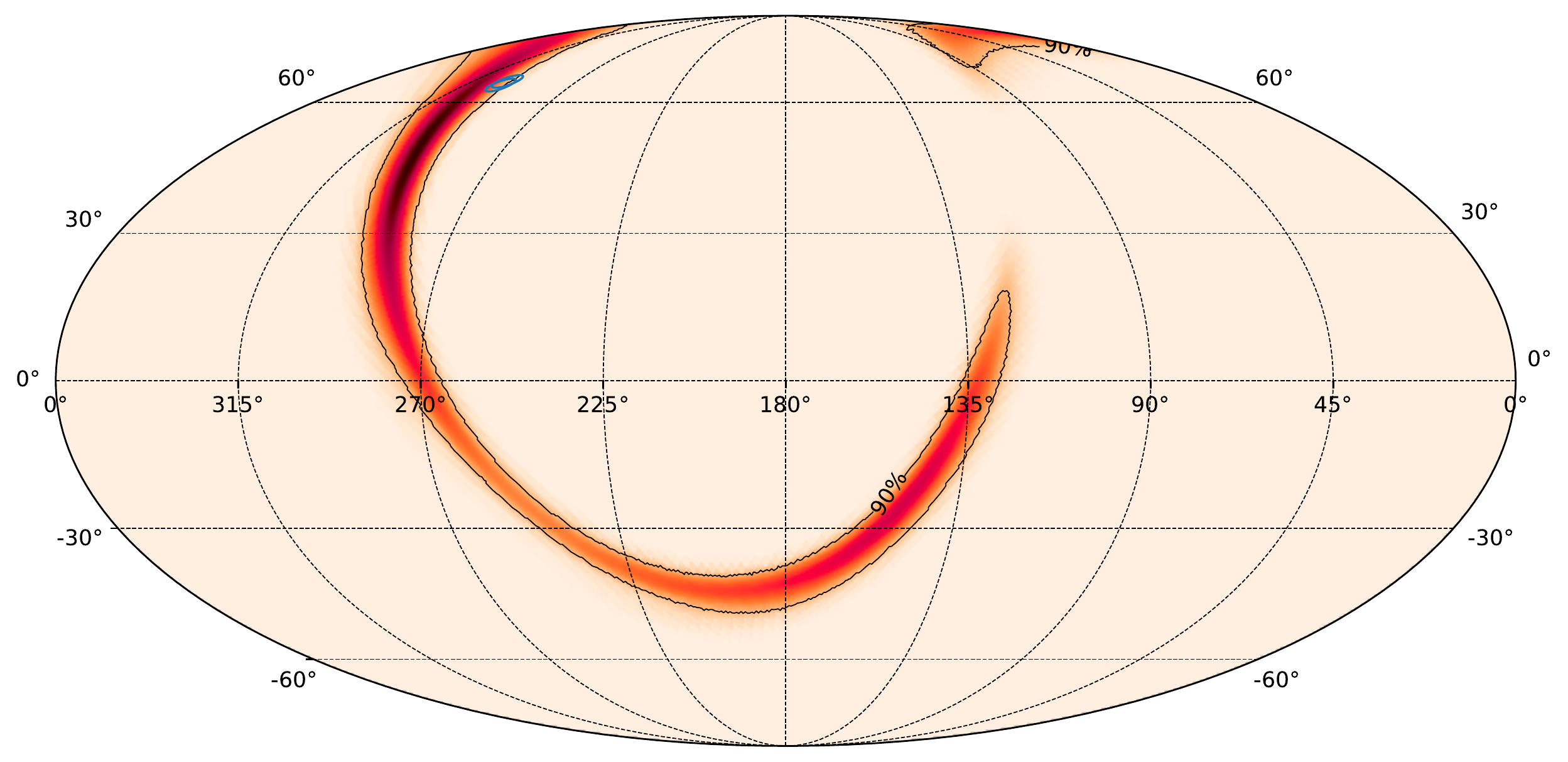}
    \caption{Spatial and temporal coincidence of VERITAS observations (blue circles) with 90\% localization (black countours) of LIGO BNS candidate C1 (2015-10-12T02\_40\_22).}
    \label{fig:C1}
\end{figure*}

\begin{figure*}[h!]
    \centering
    \includegraphics[width=\textwidth]{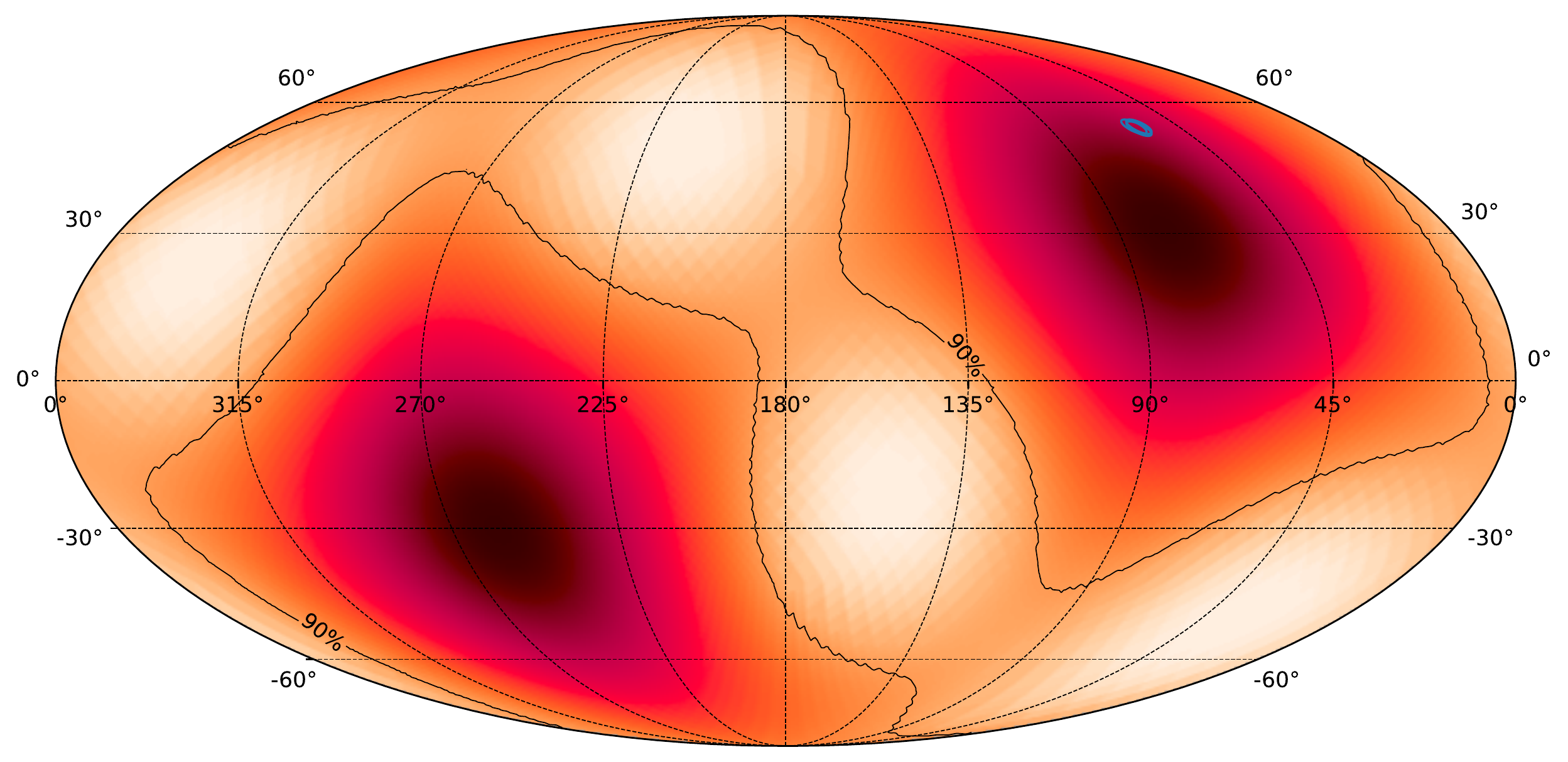}
    \caption{Spatial and temporal coincidence of VERITAS observations (blue circles) with 90\% localization (black countours) of LIGO BNS candidate C2 (2015-10-24T09\_03\_52).}
    \label{fig:C2}
\end{figure*}

\begin{figure*}[h!]
    \centering
    \includegraphics[width=\textwidth]{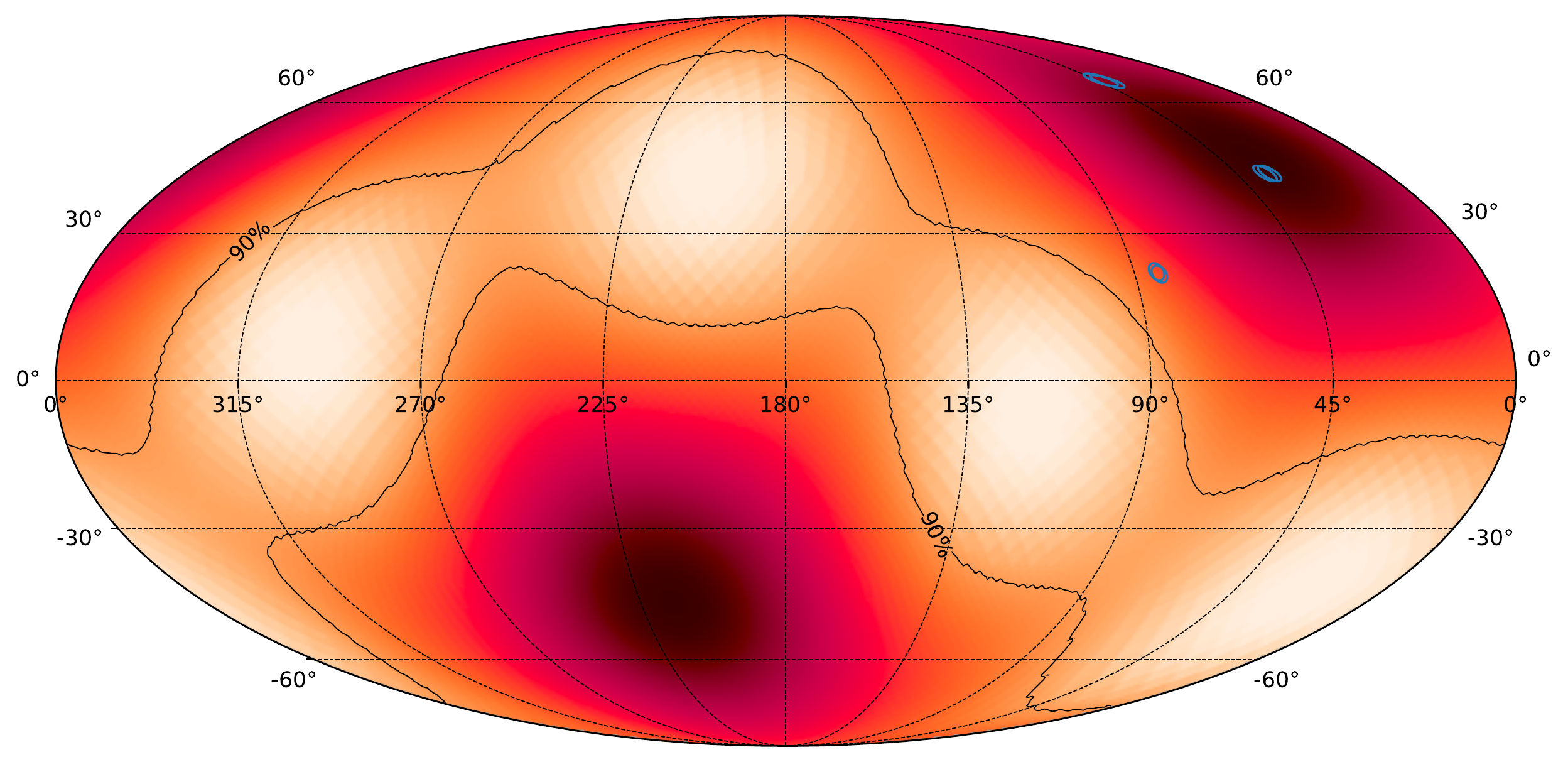}
    \caption{Spatial and temporal coincidence of VERITAS observations (blue circles) with 90\% localization (black countours) of LIGO BNS candidate C3 (2015-11-17T06\_34\_02).}
    \label{fig:C3}
\end{figure*}

\begin{figure*}[h!]
    \centering
    \includegraphics[width=\textwidth]{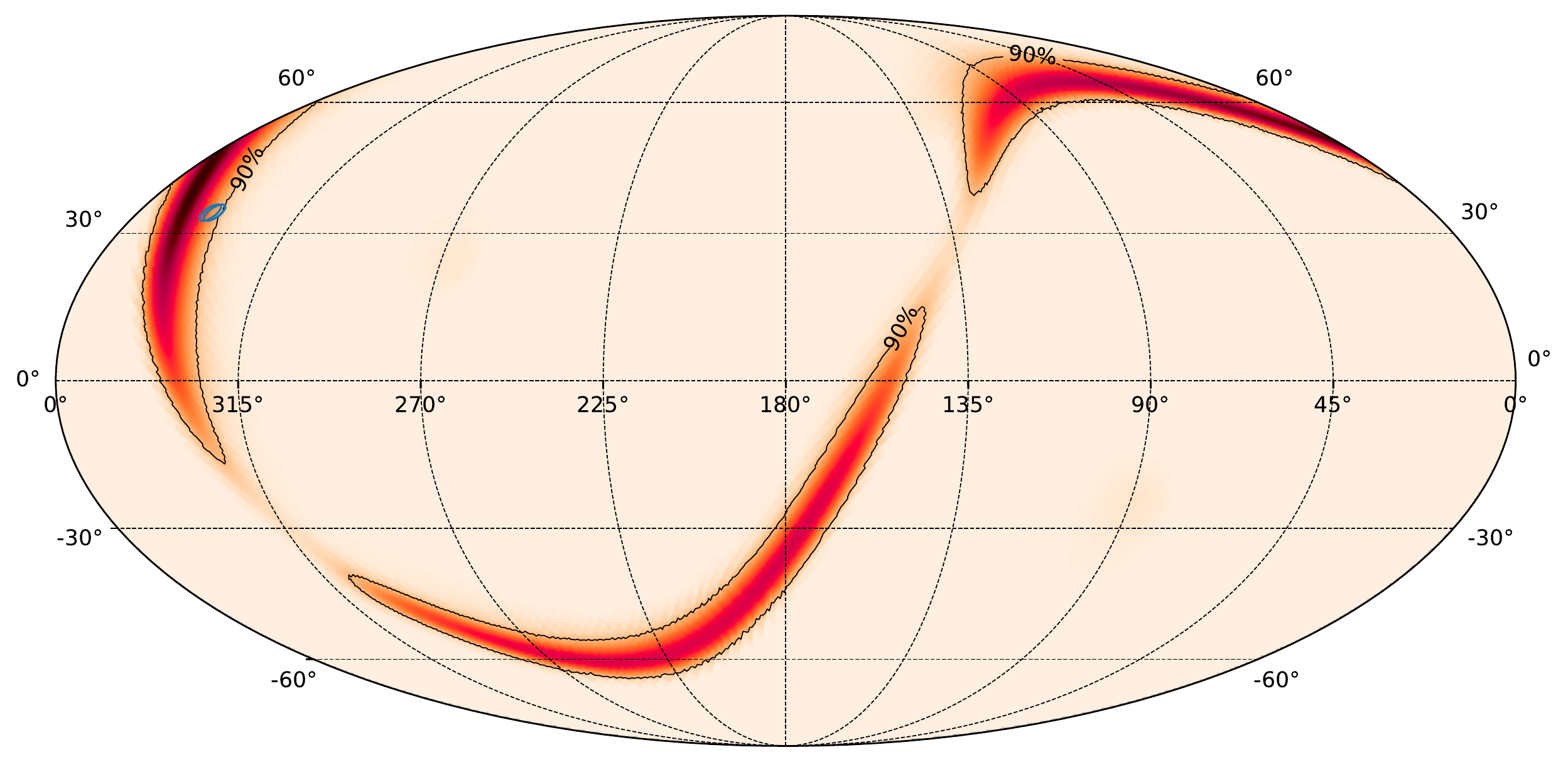}
    \caption{Spatial and temporal coincidence of VERITAS observations (blue circles) with 90\% localization (black countours) of LIGO BNS candidate C4 (2015-12-04T01\_53\_39).}
    \label{fig:C4}
\end{figure*}

\begin{figure*}[h!]
    \centering
    \includegraphics[width=\textwidth]{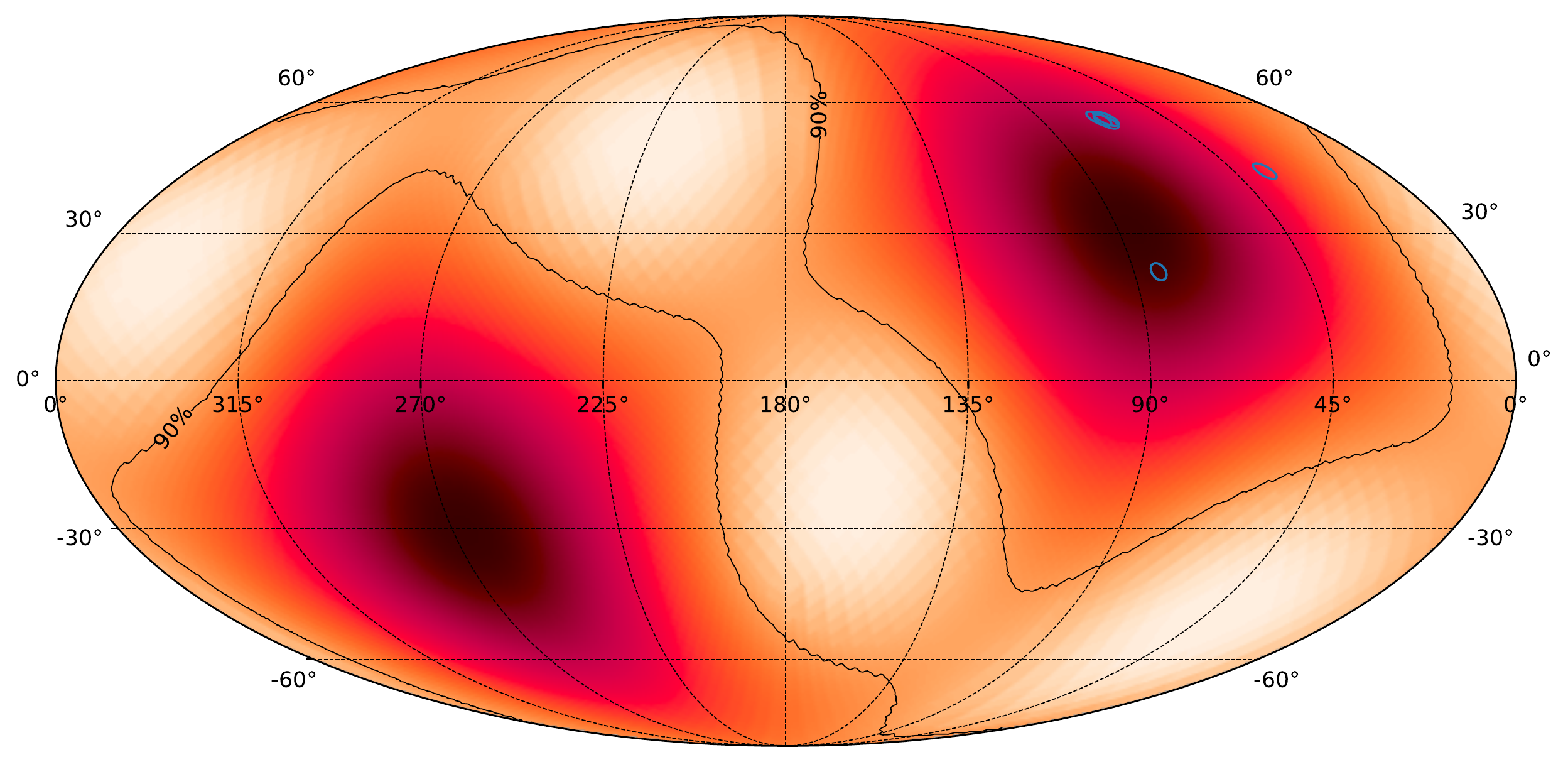}
    \caption{Spatial and temporal coincidence of VERITAS observations (blue circles) with 90\% localization (black countours) of LIGO BNS candidate C5 (2015-12-06T06\_50\_38).}
    \label{fig:C5}
\end{figure*}

\begin{figure*}[h!]
    \centering
    \includegraphics[width=\textwidth]{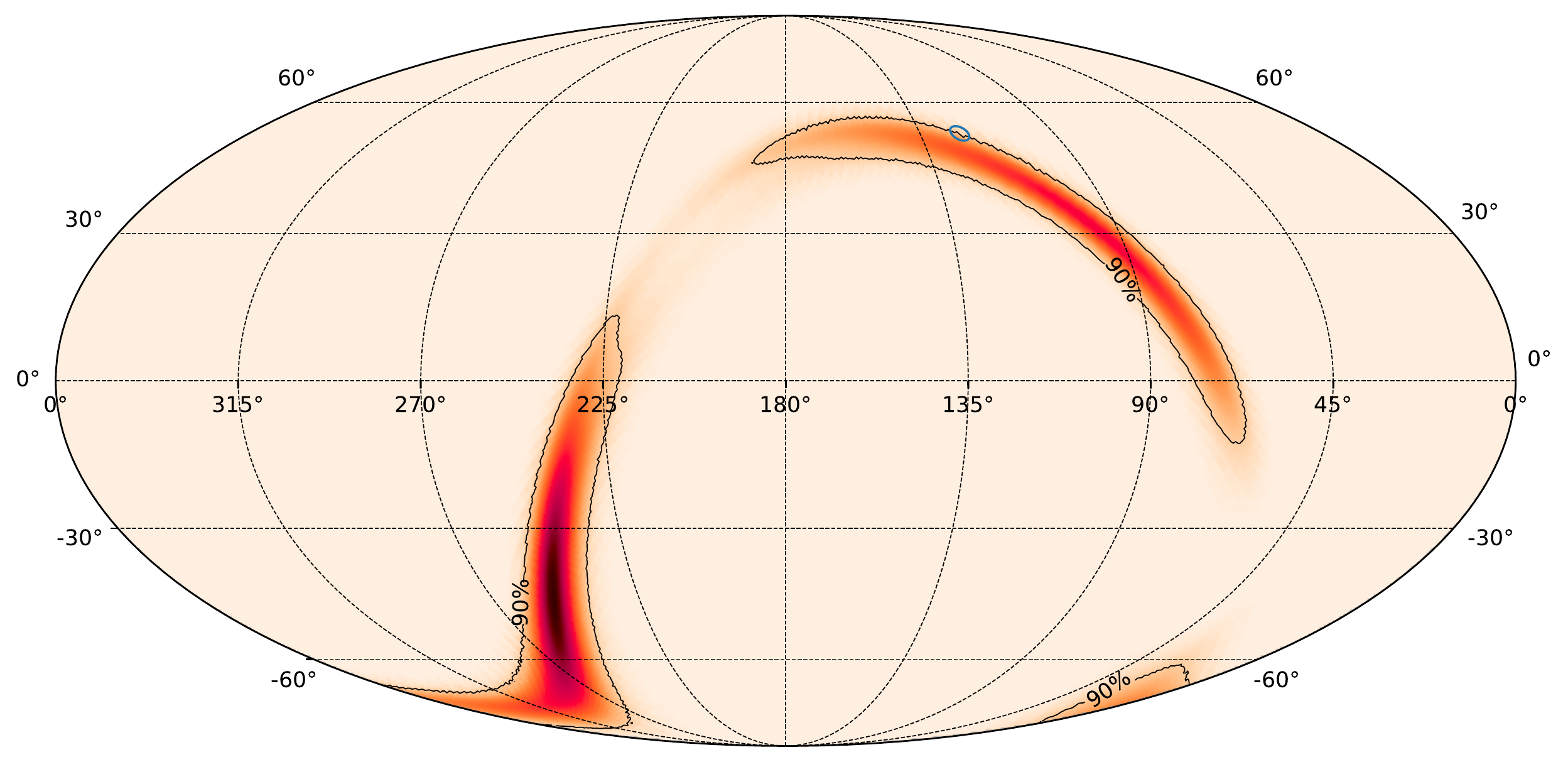}
    \caption{Spatial and temporal coincidence of VERITAS observations (blue circles) with 90\% localization (black countours) of LIGO BNS candidate C6 (2015-12-09T07\_25\_24).}
    \label{fig:C6}
\end{figure*}

\begin{figure*}[h!]
    \centering
    \includegraphics[width=\textwidth]{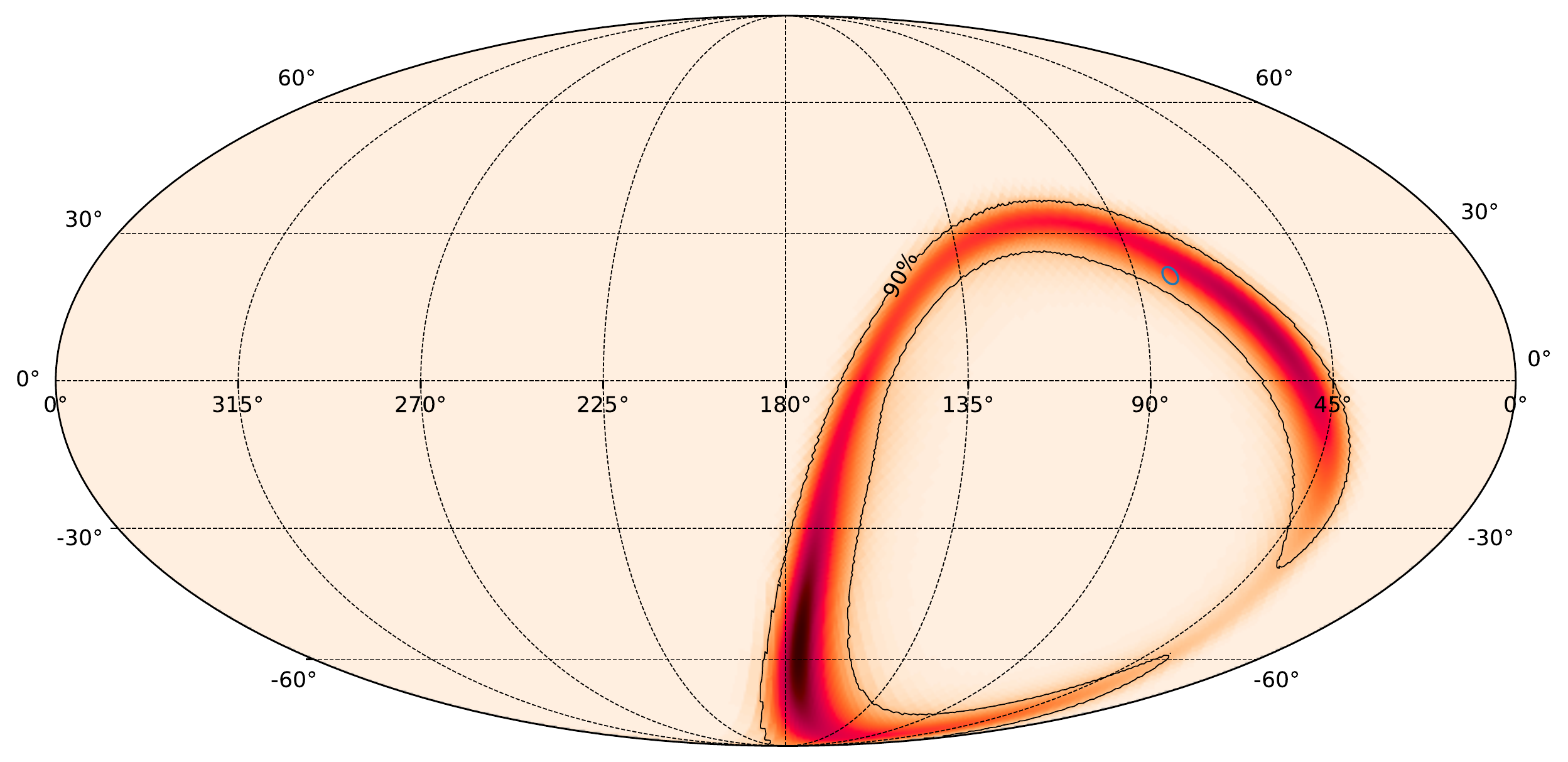}
    \caption{Spatial and temporal coincidence of VERITAS observations (blue circles) with 90\% localization (black countours) of LIGO BNS candidate C7 (2016-01-02T02\_47\_29).}
    \label{fig:C7}
\end{figure*}

\end{document}